\documentclass[fleqn,10pt]{wlscirep}
\usepackage[utf8]{inputenc}
\usepackage[T1]{fontenc}

\usepackage{caption}
\usepackage{subcaption}
\usepackage{xcolor}

\newcommand*\mywidebar[1]{%
  \hbox{%
    \vbox{%
      \hrule height 0.5pt 
      \kern0.5ex
      \hbox{%
        \kern-0.1em
        \ensuremath{#1}%
        \kern-0.1em
      }%
    }%
  }%
} 

\title{Soli-enabled Noncontact Heart Rate Detection for Sleep and Meditation Tracking}

\author[1,*]{Luzhou Xu}
\author[2]{Jaime Lien}
\author[2]{Haiguang Li}
\author[2]{Nicholas Gillian}
\author[2]{Rajeev Nongpiur}
\author[2]{Jihan Li}
\author[2]{Qian Zhang}
\author[2]{Jian Cui}
\author[2]{David Jorgensen }
\author[2]{Adam Bernstein}
\author[2]{Lauren Bedal}
\author[2]{Eiji Hayashi}
\author[2]{Jin Yamanaka}
\author[2]{Alex Lee}
\author[2]{Jian Wang}
\author[2]{D Shin}
\author[2]{Ivan Poupyrev}
\author[3]{Trausti Thormundsson}
\author[2]{Anupam Pathak}
\author[4]{Shwetak Patel}

\affil[1]{Google LLC, 6420 Sequence Drive, San Diego, CA 92121, USA}
\affil[2]{Google LLC, 1600 Amphitheatre Parkway Mountain View, CA 94043, USA}
\affil[3]{Google LLC, 19510 Jamboree Rd, Irvine, CA 92612, USA}
\affil[4]{Google LLC, 601 North 34st Street, Seattle, WA 98103, USA}
\affil[*]{email: lzxu@google.com}

\begin{abstract}

Heart rate (HR) is a crucial physiological signal that can be used to monitor health and fitness. Traditional methods for measuring HR require wearable devices, which can be inconvenient or uncomfortable, especially during sleep and meditation. Noncontact HR detection methods employing microwave radar can be a promising alternative. However, the existing approaches in the literature usually use high-gain antennas and require the sensor to face the user’s chest or back, making them difficult to integrate into a portable device and unsuitable for sleep and meditation tracking applications. This study presents a novel approach for noncontact HR detection using a miniaturized Soli radar chip embedded in a portable device (Google Nest Hub). The chip has a $6.5 \mbox{ mm} \times 5 \mbox{ mm} \times 0.9 \mbox{ mm}$ dimension and can be easily integrated into various devices. The proposed approach utilizes advanced signal processing and machine learning techniques to extract HRs from radar signals. The approach is validated on a sleep dataset (62 users, 498 hours) and a meditation dataset (114 users, 1131 minutes). The approach achieves a mean absolute error (MAE) of $1.69$ bpm and a mean absolute percentage error (MAPE) of $2.67\%$ on the sleep dataset. On the meditation dataset, the approach achieves an MAE of $1.05$ bpm and a MAPE of $1.56\%$. The recall rates for the two datasets are $88.53\%$ and $98.16\%$, respectively. This study represents the first application of the noncontact HR detection technology to sleep and meditation tracking, offering a promising alternative to wearable devices for HR monitoring during sleep and meditation.

\end{abstract}
\begin{document}
\newcommand{\re}{{\rm Re} \, }
\newcommand{\e}{{\rm E} \, }
\newcommand{\p}{{\rm P} \, }
\newcommand{\cn}{{\cal CN} \, }
\newcommand{\n}{{\cal N} \, }
\newcommand{\ba}{\begin{array}}
\newcommand{\ea}{\end{array}}
\newcommand{\be}{\begin{displaymath}}
\newcommand{\ee}{\end{displaymath}}
\newcommand{\ben}{\begin{equation}}
\newcommand{\een}{\end{equation}}
\newcommand{\bena}{\begin{eqnarray}}
\newcommand{\eena}{\end{eqnarray}}
\newcommand{\beqa}{\begin{eqnarray*}}
\newcommand{\enqa}{\end{eqnarray*}}
\newcommand{\f}{\frac}
\newcommand{\bc}{\begin{center}}
\newcommand{\ec}{\end{center}}
\newcommand{\bi}{\begin{itemize}}
\newcommand{\ei}{\end{itemize}}
\newcommand{\benu}{\begin{enumerate}}
\newcommand{\eenu}{\end{enumerate}}
\newcommand{\bdes}{\begin{description}}
\newcommand{\edes}{\end{description}}
\newcommand{\bt}{\begin{tabular}}
\newcommand{\et}{\end{tabular}}
\newcommand{\vs}{\vspace}
\newcommand{\hs}{\hspace}
\newcommand{\sort}{\rm sort \,}

\newcommand \thetabf{{\mbox{\boldmath$\theta$\unboldmath}}}
\newcommand{\Phibf}{\mbox{${\bf \Phi}$}}
\newcommand{\Psibf}{\mbox{${\bf \Psi}$}}
\newcommand \alphabf{\mbox{\boldmath$\alpha$\unboldmath}}
\newcommand \betabf{\mbox{\boldmath$\beta$\unboldmath}}
\newcommand \gammabf{\mbox{\boldmath$\gamma$\unboldmath}}
\newcommand \deltabf{\mbox{\boldmath$\delta$\unboldmath}}
\newcommand \epsilonbf{\mbox{\boldmath$\epsilon$\unboldmath}}
\newcommand \zetabf{\mbox{\boldmath$\zeta$\unboldmath}}
\newcommand \etabf{\mbox{\boldmath$\eta$\unboldmath}}
\newcommand \iotabf{\mbox{\boldmath$\iota$\unboldmath}}
\newcommand \kappabf{\mbox{\boldmath$\kappa$\unboldmath}}
\newcommand \lambdabf{\mbox{\boldmath$\lambda$\unboldmath}}
\newcommand \mubf{\mbox{\boldmath$\mu$\unboldmath}}
\newcommand \nubf{\mbox{\boldmath$\nu$\unboldmath}}
\newcommand \xibf{\mbox{\boldmath$\xi$\unboldmath}}
\newcommand \pibf{\mbox{\boldmath$\pi$\unboldmath}}
\newcommand \rhobf{\mbox{\boldmath$\rho$\unboldmath}}
\newcommand \sigmabf{\mbox{\boldmath$\sigma$\unboldmath}}
\newcommand \taubf{\mbox{\boldmath$\tau$\unboldmath}}
\newcommand \upsilonbf{\mbox{\boldmath$\upsilon$\unboldmath}}
\newcommand \phibf{\mbox{\boldmath$\phi$\unboldmath}}
\newcommand \varphibf{\mbox{\boldmath$\varphi$\unboldmath}}
\newcommand \chibf{\mbox{\boldmath$\chi$\unboldmath}}
\newcommand \psibf{\mbox{\boldmath$\psi$\unboldmath}}
\newcommand \omegabf{\mbox{\boldmath$\omega$\unboldmath}}
\newcommand \Sigmabf{\hbox{$\bf \Sigma$}}
\newcommand \Upsilonbf{\hbox{$\bf \Upsilon$}}
\newcommand \Omegabf{\hbox{$\bf \Omega$}}
\newcommand \Deltabf{\hbox{$\bf \Delta$}}
\newcommand \Gammabf{\hbox{$\bf \Gamma$}}
\newcommand \Thetabf{\hbox{$\bf \Theta$}}
\newcommand \Lambdabf{\hbox{$\bf \Lambda$}}
\newcommand \Xibf{\hbox{\bf$\Xi$}}
\newcommand \Pibf{\hbox{\bf$\Pi$}}
\newcommand \abf{{\bf a}}
\newcommand \bbf{{\bf b}}
\newcommand \cbf{{\bf c}}
\newcommand \dbf{{\bf d}}
\newcommand \ebf{{\bf e}}
\newcommand \fbf{{\bf f}}
\newcommand \gbf{{\bf g}}
\newcommand \hbf{{\bf h}}
\newcommand \ibf{{\bf i}}
\newcommand \jbf{{\bf j}}
\newcommand \kbf{{\bf k}}
\newcommand \lbf{{\bf l}}
\newcommand \mbf{{\bf m}}
\newcommand \nbf{{\bf n}}
\newcommand \obf{{\bf o}}
\newcommand \pbf{{\bf p}}
\newcommand \qbf{{\bf q}}
\newcommand \rbf{{\bf r}}
\newcommand \sbf{{\bf s}}
\newcommand \tbf{{\bf t}}
\newcommand \ubf{{\bf u}}
\newcommand \vbf{{\bf v}}
\newcommand \wbf{{\bf w}}
\newcommand \xbf{{\bf x}}
\newcommand \ybf{{\bf y}}
\newcommand \zbf{{\bf z}}
\newcommand \rbfa{{\bf r}}
\newcommand \xbfa{{\bf x}}
\newcommand \ybfa{{\bf y}}
\newcommand \Abf{{\bf A}}
\newcommand \Bbf{{\bf B}}
\newcommand \Cbf{{\bf C}}
\newcommand \Dbf{{\bf D}}
\newcommand \Ebf{{\bf E}}
\newcommand \Fbf{{\bf F}}
\newcommand \Gbf{{\bf G}}
\newcommand \Hbf{{\bf H}}
\newcommand \Ibf{{\bf I}}
\newcommand \Jbf{{\bf J}}
\newcommand \Kbf{{\bf K}}
\newcommand \Lbf{{\bf L}}
\newcommand \Mbf{{\bf M}}
\newcommand \Nbf{{\bf N}}
\newcommand \Obf{{\bf O}}
\newcommand \Pbf{{\bf P}}
\newcommand \Qbf{{\bf Q}}
\newcommand \Rbf{{\bf R}}
\newcommand \Sbf{{\bf S}}
\newcommand \Tbf{{\bf T}}
\newcommand \Ubf{{\bf U}}
\newcommand \Vbf{{\bf V}}
\newcommand \Wbf{{\bf W}}
\newcommand \Xbf{{\bf X}}
\newcommand \Ybf{{\bf Y}}
\newcommand \Zbf{{\bf Z}}
\newcommand \Omegabbf{{\bf \Omega}}
\newcommand \Rssbf{{\bf R_{ss}}}
\newcommand \Ryybf{{\bf R_{yy}}}
\newcommand \Cset{{\cal C}}
\newcommand \Rset{{\cal R}}
\newcommand \Zset{{\cal Z}}
\newcommand{\otheta}{\stackrel{\circ}{\theta}}
\newcommand{\defeq}{\stackrel{\bigtriangleup}{=}}
\newcommand{\oabf}{{\bf \breve{a}}}
\newcommand{\odbf}{{\bf \breve{d}}}
\newcommand{\oDbf}{{\bf \breve{D}}}
\newcommand{\oAbf}{{\bf \breve{A}}}
\renewcommand \vec{{\mbox{vec}}}
\newcommand{\Acalbf}{\bf {\cal A}}
\newcommand{\calZbf}{\mbox{\boldmath $\cal Z$}}
\newcommand{\feop}{\hfill \rule{2mm}{2mm} \\}
\newtheorem{theorem}{Theorem}[section]

\newcommand{\Rnum}{{\mathbb R}}
\newcommand{\Cnum}{{\mathbb C}}
\newcommand{\Znum}{{\mathbb Z}}

\newcommand{\Ccal}{{\cal C}}
\newcommand{\Dcal}{{\cal D}}
\newcommand{\Hcal}{{\cal H}}
\newcommand{\Ocal}{{\cal O}}
\newcommand{\Rcal}{{\cal R}}
\newcommand{\Zcal}{{\cal Z}}
\newcommand{\Xcal}{{\cal X}}
\newcommand{\zzbf}{{\bf 0}}
\newcommand{\zebf}{{\bf 0}}

\newcommand{\eop}{\hfill $\Box$}

\newcommand{\gss}{\mathop{}\limits}
\newcommand{\gs}{\mathop{\gss_<^>}\limits}

\newcommand{\circlambda}{\mbox{$\Lambda$
             \kern-.85em\raise1.5ex
             \hbox{$\scriptstyle{\circ}$}}\,}

\newcommand{\tr}{\mathop{\rm tr}}
\newcommand{\var}{\mathop{\rm var}}
\newcommand{\cov}{\mathop{\rm cov}}
\newcommand{\diag}{\mathop{\rm diag}}
\def\rank{\mathop{\rm rank}\nolimits}
\newcommand{\ra}{\rightarrow}
\newcommand{\ul}{\underline}
\def\Pr{\mathop{\rm Pr}}
\def\Re{\mathop{\rm Re}}
\def\Im{\mathop{\rm Im}}

\def\submbox#1{_{\mbox{\footnotesize #1}}}
\def\supmbox#1{^{\mbox{\footnotesize #1}}}

%
\newtheorem{Theorem}{Theorem}[section]
\newtheorem{Definition}[Theorem]{Definition}
\newtheorem{Proposition}[Theorem]{Proposition}
\newtheorem{Lemma}[Theorem]{Lemma}
\newtheorem{Corollary}[Theorem]{Corollary}
%
%
\newcommand{\ThmRef}[1]{\ref{thm:#1}}
\newcommand{\ThmLabel}[1]{\label{thm:#1}}
\newcommand{\DefRef}[1]{\ref{def:#1}}
\newcommand{\DefLabel}[1]{\label{def:#1}}
\newcommand{\PropRef}[1]{\ref{prop:#1}}
\newcommand{\PropLabel}[1]{\label{prop:#1}}
\newcommand{\LemRef}[1]{\ref{lem:#1}}
\newcommand{\LemLabel}[1]{\label{lem:#1}}
%

\flushbottom
\maketitle
%
%
\thispagestyle{empty}


\section*{Introduction}

Sleep is essential for overall well-being, having a significant impact on physical and mental health. Research indicates that poor sleep quality and inadequate sleep can lead to various health issues, including obesity, diabetes, cardiovascular disease, and depression \cite{CDC2021}. Additionally, meditation has been found to have multiple benefits, such as reducing stress, anxiety, and depression and enhancing attention and cognitive function \cite{Tang2015}. Therefore, accurately tracking sleep and meditation patterns is crucial for individuals to improve their practice and maintain optimal health and wellness \cite{Kratz2020}. 

{The integration of smart beds/mattresses, equipped with advanced sensing technologies such as pressure sensors \cite{SpillmanJr2004, LaurinoArcarisiCarbonaro2020}, presents a promising avenue for systematic, preventive, personalized, and non-invasive assessment of sleep quality. These technologies enable the measurement of critical parameters like respiration rate (RR) and heart rate (HR). However, their limitations, including bulkiness, high cost, and immobility, pose challenges to widespread adoption and deployment in diverse settings.} 
On the contrary, wearable devices like smartwatches have gained popularity for sleep and meditation tracking. They can collect data on movement and HR using sensors like accelerometers and Photoplethysmography (PPG) \cite{Olsen2020}. Nevertheless, wearing a device during sleep or meditation can be uncomfortable and inconvenient, potentially leading to incomplete or inaccurate data \cite{Olsen2020}. 
{Remote photoplethysmography (rPPG) \cite{Poh2010,Haan2014,BoccignoneConteCuculo2020,BaeBorac2022} has also emerged as a compelling alternative to wearable devices. It offers valuable insights, including heart rate (HR) measurement, by detecting changes in reflected light intensity from a user's skin. However, rPPG's applicability is limited by its requirement to "see" the user's skin, making it unsuitable for comprehensive sleep and meditation tracking due to constraints related to varying light conditions, sleep postures, and concerns regarding user privacy.}

Portable devices like the Google Nest Hub \cite{google_nest_hub_sleep_sensing}, equipped with microwave radar technology, introduce an innovative approach to monitoring sleep and meditation quality. These devices eliminate the discomfort and inconvenience associated with wearing tracking devices by providing valuable insights through motion tracking and respiratory rate (RR) monitoring \cite{lin1975microwave}. In comparison to their rPPG \cite{Poh2010,Haan2014, BoccignoneConteCuculo2020,BaeBorac2022} counterparts, radar systems offer several advantages, including enhanced privacy protection, independence from light conditions, and reduced power consumption. They can also penetrate clothing and blankets, and are suitable for various postures during sleep and meditation. Yet, it's important to acknowledge that current portable devices, including the Google Nest Hub, lack the ability to monitor heart rate (HR), a crucial biometric for comprehensive sleep and meditation tracking, due to existing technical challenges. The implementation of HR measurement in portable devices is an area of great interest.

\begin{figure}
     \centering
     \begin{subfigure}[b]{0.35\textwidth}
         \centering
         \includegraphics[width=\textwidth]{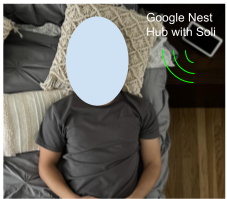}
         \caption{}
         \label{fig1a}
     \end{subfigure}
     \begin{subfigure}[b]{0.5\textwidth}
         \centering
         \includegraphics[width=\textwidth]{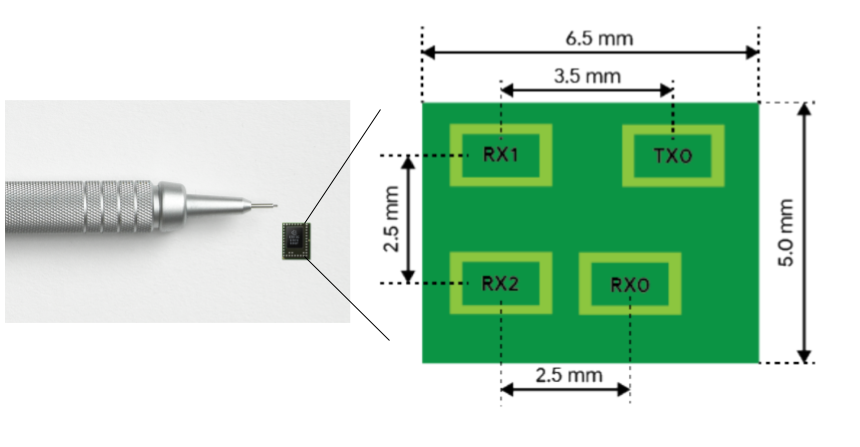}
         \caption{}
         \label{fig1b}
     \end{subfigure}
     \medskip
     \hfill
     \begin{subfigure}[b]{0.95\textwidth}
         \centering
         \includegraphics[width=\textwidth]{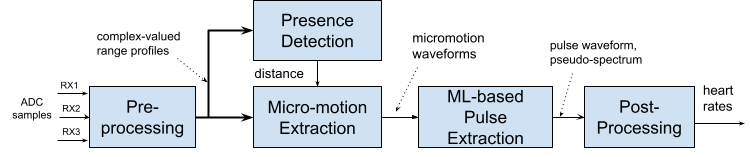}
         \caption{}
         \label{fig1c}
     \end{subfigure}
     \medskip
      \hfill
     \begin{subfigure}[b]{0.95\textwidth}
         \centering
         \includegraphics[width=\textwidth]{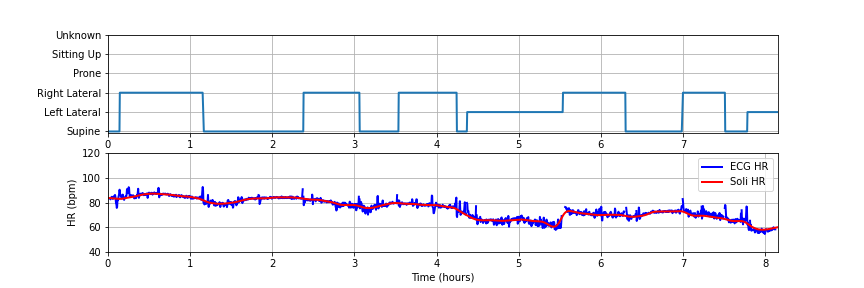}
         \caption{}
         \label{fig1d}
     \end{subfigure}
        \caption{(a) A Google Nest Hub monitors a user's HR contactlessly on a bedside table. The device integrates a radar chip called Soli, operating at the 60 GHz frequency band. (b) The Soli radar chip is 6.5 mm x 5 mm x 0.9 mm and comprises one transit antenna and 3 receive antennas on the package. (c) The proposed HR detection approach includes three signal processing blocks to detect the user's presence and extract micro-motion waveforms, a machine learning block to extract pulse waveform and related pseudo-spectrum, and a post-processing block to smooth HRs. (d) An overnight sleep example is shown in this plot. The upper plot illustrates the user's sleep position, while the lower plot shows the HRs estimated by the proposed method (i.e., Soli HR) and the ECG HRs as the ground truth. The Soli HR can accurately track slow HR fluctuations over time under various sleep positions.
}
        \label{fig1}
\end{figure}

\section*{Related Work}

Noncontact HR detection using microwave radar \cite{li2009radar,LiLubeckeVictor2013,LiPengHuang2017,Mercuri2019,park2021preclinical} represents a promising technology to integrate HR monitoring into portable sleep- and meditation-tracking devices. This approach emits electromagnetic waves toward the target subject. It measures the reflected signals, which exhibit periodic variations related to the HR due to changes in blood flow with each heartbeat. HR information can be extracted from the reflected signals using advanced signal processing (SP) and machine learning (ML) techniques without physical contact. Although significant efforts have been dedicated to this field over the past few decades, two challenges still need to be solved for using radar in portable sleep and meditation-tracking devices. 

Firstly, most existing systems rely on high-gain antennas to amplify signal-to-noise ratio (SNR) and diminish environmental interference by directing the radiating energy onto the user’s body. Nonetheless, this method necessitates larger antenna sizes and produces a narrower field of view (FOV), limiting the user’s positioning to typically being right in front of the radar sensor.

Secondly, the weak micro-motion of the heartbeat is often overshadowed by respiration motion and random body motion, which are typically much stronger than the heartbeat by one or two orders of magnitude. In the frequency domain, the weak heartbeat signal can be easily masked by the harmonic components and sidelobes of these two interfering motions. Thus, traditional spectral analysis techniques, such as Fourier transform (FT) and band-pass filter (BPF) \cite{park2021preclinical}, fail to perform satisfactorily. Various methods have been developed to address this issue, including ensemble empirical mode decomposition (EEMD) \cite{MotinKarmakar2016}, wavelet \cite{TariqShiraz2011}, RELAX \cite{LiJingLi2010}, Kalman filter \cite{ArsalanSantraWIll2020}, independent component analysis (ICA) \cite{MuramatsuYamamoto2022}, and ML \cite{IyerZhaoPrabhakar2022}. However, the authors of these papers have primarily focused on cases where the radar sensor is directed straight at the user's chest or back. Detecting HR from the sides presents an even more significant challenge due to the smaller magnitude of heartbeat micro-motion and the obstruction of the arms and shoulders. In fact, a recent study \cite{IyerZhaoPrabhakar2022} discovered that ``{\textit{ the left and right orientations are unsuitable for monitoring HR and RR}}'' when examining the HR performance for different user orientations. Unfortunately, users may assume various poses during sleep. The most common sleep position, supine, leads to the side orientation, making it difficult to acquire reliable HRs  using a portable device at the bedside. More advanced techniques are required to overcome this challenge.

This article introduces an innovative approach for continuously monitoring a user's HR during sleep and meditation, utilizing a Google Nest Hub placed on a bedside table (Figure \ref{fig1a}). The Google Nest Hub incorporates a miniaturized radar chip named Soli, which operates at the $60$ GHz frequency band and includes a transit antenna and 3 receive antennas, all on the package, with a small dimension of $6.5 \mbox{ mm} \times 5 \mbox{ mm} \times 0.9 \mbox{ mm}$ (Figure \ref{fig1b}). We propose a combined SP and ML approach to address the challenges of weak heartbeat micro-motion and interference from respiration and random body motion (as depicted in Figure \ref{fig1c}). The approach involves using three SP blocks to detect the user's presence and extract micro-motion waveforms from different body parts. Then, an ML-based pulse extraction block is used to extract the heart pulse waveform and associated pseudo-spectrum from the micro-motion waveforms, from which we detect the HR. Finally, a post-processing block is employed to smooth the HR sequence. 

We have tested the proposed method on a sleep dataset (62 users, 498 hours) and a meditation dataset (114 users, 1131 minutes), which results in mean-absolute-error (MAE) of $1.69$ bpm (beat-per-minute) and $1.05$ bpm,  mean-absolute-percentage-error (MAPE) of $2.67\%$ and $2.0\%$, and recall rates of  $88.53\%$ and $98.16\%$, respectively. We provide a sleep-tracking example in Figure \ref{fig1d}, which shows the overnight Soli HR sequence compared to its electrocardiogram (ECG) counterpart and the user's sleep position. Soli's HR estimates precisely track gradual HR changes across different sleep positions, and they closely match with ECG HRs.

To the best of our knowledge, this article represents a significant contribution to three areas of research: (1) the use of a miniaturized radar chip in a commercial portable device for noncontact HR detection; (2) the study of noncontact HR detection for sleep and meditation tracking, including challenging sleep positions where the radar sensor views users from their sides; and (3) the evaluation of this technology on large-scale datasets, including 62 users' overnight sleep data and 112 users' meditation data under various practical conditions. Overall, it introduces the novel application of noncontact HR detection technology to sleep and meditation tracking, providing a promising solution for noncontact HR monitoring in these domains.

\begin{figure}
     \centering
     \begin{subfigure}[b]{0.45\textwidth}
         \centering
         \includegraphics[width=\textwidth]{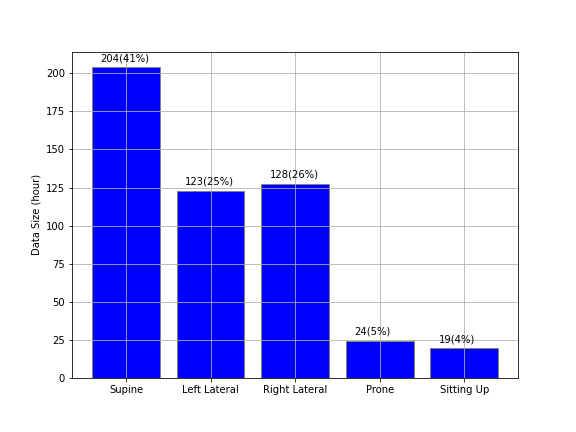}
         \caption{}
         \label{fig_csss_dist_a}
     \end{subfigure}
     \begin{subfigure}[b]{0.45\textwidth}
         \centering
         \includegraphics[width=\textwidth]{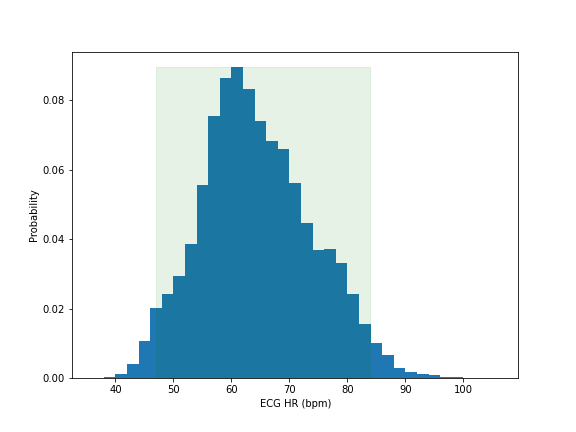}
         \caption{}
         \label{fig_csss_dist_b}
     \end{subfigure}
     \caption{(a) The distribution of sleep data for different sleep positions: supine, left lateral, right lateral, prone, and sitting-up, which account for $41\%$, $25\%$, $26\%$, $5\%$, and $4\%$ of the total data, respectively. (b) The ECG HR histogram ranges from 40 bpm to 100 bpm. 
}
     \label{fig_csss_dist}
\end{figure}

\section*{Results}

Below, we present the HR results on the sleep and meditation datasets.

\subsection*{Sleep Tracking}

To collect sleep data, we used a Google Nest Hub placed on a bedside table (refer to Figure \ref{fig1a}) to record Soli radar data. The distance between the Soli sensor and the user's torso varied from 0.3 to 1.5 meters. We also attached an ECG sensor to the user's body to record ECG signals as a reference. For each session, a user was brought in to sleep overnight, ranging from 6 to 9 hours. We had a proctor present to record the user's sleep positions and events (such as apnea, hypopnea, and arousal). 

\begin{figure}
     \centering
     \begin{subfigure}[b]{0.45\textwidth}
         \centering
         \includegraphics[width=\textwidth]{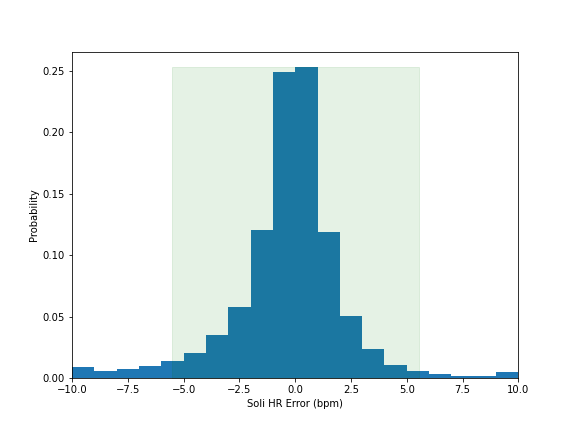}
         \caption{}
         \label{fig_csss_hr_err_hist}
     \end{subfigure}
     \begin{subfigure}[b]{0.45\textwidth}
         \centering
         \includegraphics[width=\textwidth]{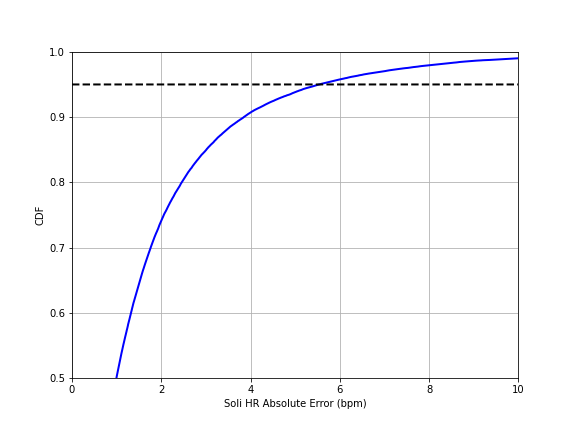}
         \caption{}
         \label{fig_csss_hr_err_cdf}
     \end{subfigure}
     \caption{(a) The HR error histogram of the proposed approach during sleep, and (b) the related AE CDF. The histogram indicates that the majority of HR errors are small. The AE cumulative density function shows that the 95th percentile AE is less than 5.50 bpm.}
     \label{fig_csss_hr_err_dist}
\end{figure}

A total of $232$ sessions of data have been collected. Of these, $66$ sessions do not have ECG data files, $41$ sessions have a timestamp misalignment issue between Soli and ECG data, and 3 sessions have poor-quality ECG signals. After removing these sessions, $122$ sessions with valid ECG data were identified. These sessions correspond to $122$ unique users and a total of 975 hours. We randomly partitioned the $122$ sessions into two groups, $60$ sessions ($477$ hours) for ML training and $62$ sessions ($498$ hours) for performance evaluation. As shown in Figure \ref{fig_csss_dist_a}, five different sleep positions were recorded from these $62$ test sessions: supine ($204$ hours, $41\%$ of total data), left lateral ($123$ hours, $25\%$), right lateral ($128$ hours, $26\%$), prone ($24$ hours, $5\%$), and sitting-up ($19$ hours, $4\%$). The reference ECG HRs, {derived by applying FFT to the ECG waveforms}, range from $40$ to $100$ bpm, with $95\%$ of the samples falling within the range of $45$ to $85$ bpm (Figure \ref{fig_csss_dist_b}). The sleep-tracking HR performance below is based on the $62$ test sessions.

During data processing, the overnight data is segmented into 60-second overlapping samples with a step size of $15$ seconds. Each $60$-second sample is processed through SP and ML blocks, as illustrated in Figure \ref{fig1c}, to obtain an HR estimate. The post-processing block then smooths the HR estimates over time. Thus, the first HR estimate is obtained in 60 seconds and updated every $15$ seconds. The performance metrics are evaluated based on the smoothed HR estimates.

\begin{table}[ht]
\centering
\begin{tabular}{|p{0.1\linewidth}|p{0.1\linewidth}|p{0.1\linewidth}|p{0.1\linewidth}|p{0.1\linewidth}|p{0.1\linewidth}|p{0.1\linewidth}|}
\hline
\textbf{Method} & \textbf{MAE} & \textbf{AE} ($95\%$) &  \textbf{MAPE}  &  \textbf{APE} ($95\%$) & {\textbf{R$^2$}} &  \textbf{Recall}  \\
\hline
\textbf{BPF} &  $21.0$ bpm & $67.3$ bpm & $31.2\%$ & $98.5\%$ & - &$73\%$  \\  
\hline
\textbf{Proposed} & $\mathbf{1.69}$ bpm & $\mathbf{5.50}$ bpm & $\mathbf{2.67\%}$ & $\mathbf{8.49\%}$ &  {$\mathbf{89.92\%}$} 
 & $\mathbf{88.53\%}$   \\  
\hline
\end{tabular}
\caption{Sleep-tracking HR accuracies of the proposed approach and the conventional BPF method. The proposed approach achieves significantly better HR accuracy than its BPF counterpart.}
\label{tab_sleep_sense_hr_accu}
\end{table}

To provide a benchmark for comparison, we evaluate the conventional BPF approach \cite{park2021preclinical} on the sleep dataset. {This comparison with BPF, a widely accepted reference method, serves to establish a familiar baseline for evaluating the effectiveness and novelty of our approach.} The ML block in Figure \ref{fig1c} is replaced by several SP blocks, including a BPF block (with BPF passing frequency band being $40 \textup{--}200$ bpm) to extract heart pulse waveforms, an FT block to obtain the related spectrum, and a block to select the best HR over various micro-motion waveforms based on the peak-to-average ratio (PAR). As shown in Table \ref{tab_sleep_sense_hr_accu}, the BPF approach yields MAE and MAPE values of up to $21.0$ bpm and $31.2\%$, respectively. The $95$th percentile absolute error (AE) and absolute percentage error (APE) of the BPF approach are up to $67.3$ bpm and $98.5\%$, respectively. The recall rate is $73\%$, indicating that the BPF approach can not determine the HRs in $27\%$ of the samples. {Furthermore, the R-squared value, also known as the determination coefficient and defined in Equation (\ref{equ_r_square}), returns a negative value because the mean squared error (MSE) exceeds the HR variance.} These performance metrics indicate that the BPF approach failed to perform on the sleep data. 

\begin{figure}
     \centering
     \begin{subfigure}[b]{0.45\textwidth}
         \centering
         \includegraphics[width=\textwidth]{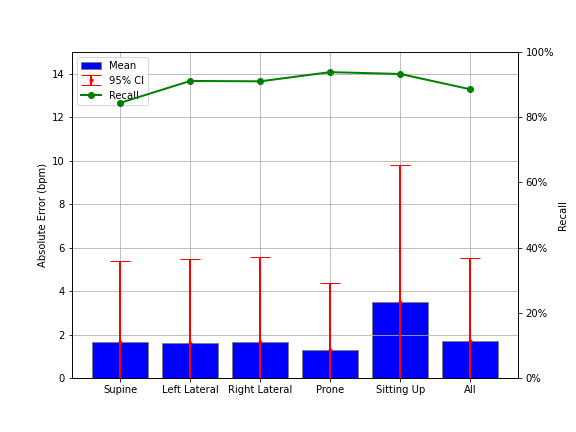}
         \caption{}
         \label{fig_csss_mae_a}
     \end{subfigure}
     \begin{subfigure}[b]{0.45\textwidth}
         \centering
         \includegraphics[width=\textwidth]{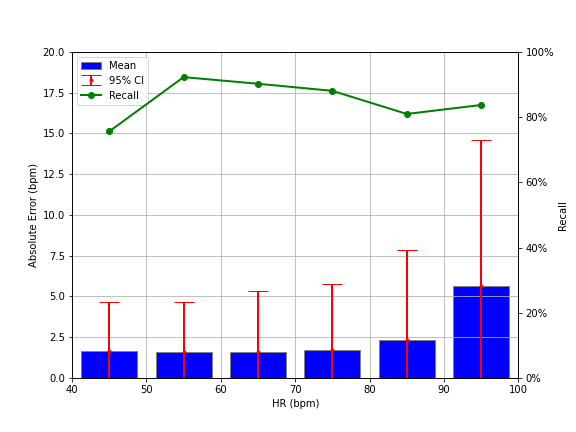}
         \caption{}
         \label{fig_csss_mae_b}
     \end{subfigure}
      \begin{subfigure}[b]{0.9\textwidth}
         \centering
         \includegraphics[width=\textwidth]{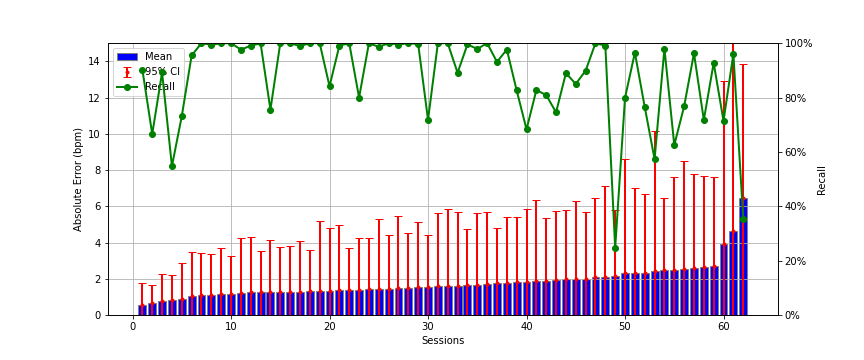}
         \caption{}
         \label{fig_csss_mae_c}
     \end{subfigure}
     \caption{(a) The HR accuracy of the proposed approach is shown for various sleep positions. The proposed approach achieves an MAE of less than $1.7$ bpm for all sleep positions except for the sitting-up pose, which is $3.48$ bpm. (b) The HR accuracy of the proposed approach is shown for various actual HR bands. The proposed approach achieves an MAE of less than $2.4$ bpm for all HR bands except for the $90\textup{--}100$ bpm band, where the MAE is $5.65$ bpm. (c) The HR accuracy of the proposed approach is shown for various sessions. The proposed approach achieves an MAE of less than $3$ bpm for $59$ out of $62$ sessions ($95.2\%$).}
     \label{fig_csss_mae}
\end{figure}

In contrast, our proposed method achieves good HR accuracy, as evidenced by its MAE and MAPE values of $1.69$ bpm and $2.67\%$, respectively. The $95$th percentile AE and APE are $5.50$ bpm and $8.49\%$, indicating that our method achieves good HR accuracy in $95\%$ of the samples. {The R-Squared value is $89.92\%$.} Moreover, the method has a recall rate of $88.53\%$, with the remaining $11.47\%$ of samples (not sessions or users) undetermined. Detailed visual representations, Figures \ref{fig_csss_hr_err_hist} and \ref{fig_csss_hr_err_cdf}, depict the distribution of HR errors through the error histogram and the corresponding AE cumulative density function (CDF). {Note that the recall rate in the sleep study is relatively low compared to its meditation counterpart, a topic to be elaborated upon in the following section.  This discrepancy can mainly be attributed to environmental factors. The sleep data collection occurred overnight in an uncontrolled environment, leading to increased body motion of participants due to the noisy surroundings and discomfort caused by ECG attachment. These factors negatively impacted both the recall rate and HR accuracy.}

The HR accuracies are further analyzed in Figures \ref{fig_csss_mae_a}, \ref{fig_csss_mae_b}, and \ref{fig_csss_mae_c} for various sleep positions, HR bands, and data sessions. In each figure, blue bars indicate MAEs, red bars indicate the $95$th percentile AE, and green curves represent recall rates. The y-axes on the left and right sides show HR errors in bpm and percentage recall rates, respectively. The results in Figure  \ref{fig_csss_mae_a} demonstrate that the four lying sleep positions (supine, left lateral, right lateral, and prone) yield similar HR accuracies with an MAE of less than 1.7 bpm and a $95$th percentile AE of less than $5.6$ bpm. The sitting-up position performs slightly worse than the lying positions, with an MAE of $3.48$ bpm and a $95$th percentile AE of $9.81$ bpm.

Similarly, in Figure \ref{fig_csss_mae_b}, we observe good HR accuracy when the true HR is between $40$ bpm and $90$ bpm, with an MAE of less than $2.4$ bpm and a $95$th percentile AE of less than $5.8$ bpm. The HR accuracy degrades slightly when the true HR is between $90$ bpm and $100$ bpm, with an MAE of $5.65$ bpm and a $95$th percentile AE of $14.60$ bpm. The lower HR accuracy for the sitting-up position and high HR bands can be traced back to the small data size available for these two categories (Figures \ref{fig_csss_dist_a} and \ref{fig_csss_dist_b}). The insufficient data size causes the ML model to be inadequately trained for these categories, leading to worse performance.

\begin{figure}
     \centering
       \begin{subfigure}[b]{0.95\textwidth}
         \centering
         \includegraphics[width=\textwidth]{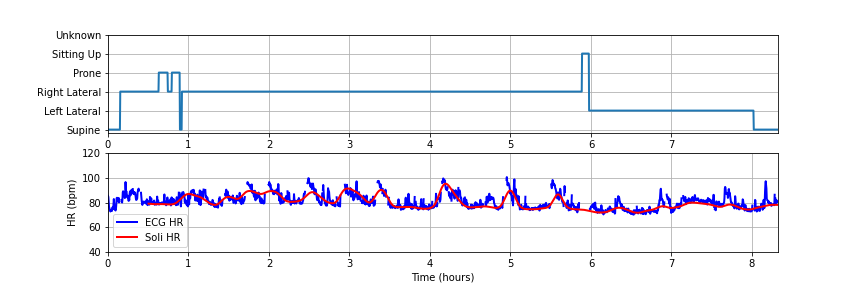}
         \caption{}
         \label{fig_csss_hr_seq_a}
     \end{subfigure}
     \begin{subfigure}[b]{0.95\textwidth}
         \centering
         \includegraphics[width=\textwidth]{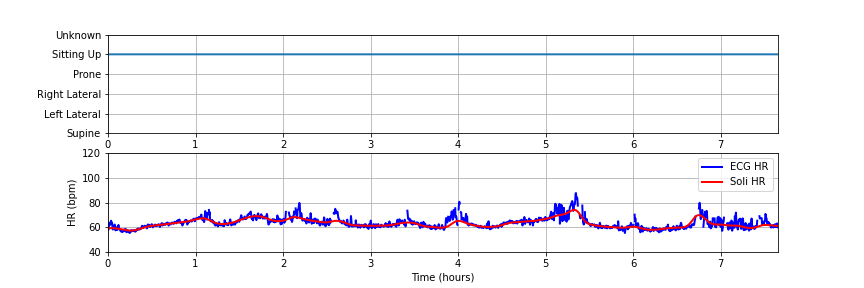}
         \caption{}
         \label{fig_csss_hr_seq_b}
     \end{subfigure}
        \caption{Overnight sleep-tracking examples using the Soli radar chip. (a) shows the oscillating HR variation tracked by the proposed approach during overnight sleep, while (b) demonstrates its capability to detect HR in a sitting-up position.
}
        \label{fig_csss_hr_seq}
\end{figure}

The HR accuracy over different sessions (i.e., users) is shown in Figure  \ref{fig_csss_mae_c}, where sessions are sorted based on their MAEs from the smallest to the largest. As per the figure, $59$ sessions ($95.2\%$) have MAEs less than $3$ bpm, $58$ sessions ($93.6\%$) have the $95$th percentile AE less than 10 bpm, and 60 sessions (which is $96.8\%$) have a recall rate greater than $50\%$. Overall, $57$ sessions ($91.9\%$) perform satisfactorily with MAE less than $3$ bpm, the $95$th percentile AE less than $10$ bpm, and a recall rate greater than $50\%$.

Figures \ref{fig1d}, \ref{fig_csss_hr_seq_a}, and \ref{fig_csss_hr_seq_b} exhibit three examples of overnight sleep tracking, with the user sleep position in the upper plot and the proposed Soli HR and ECG HR (ground truth) displayed in the lower plot. The Soli HR estimates align well with the ECG HR, accurately tracking slow HR fluctuations over time. Figure \ref{fig_csss_hr_seq_a} illustrates Soli's ability to track oscillating HR variations during overnight sleep, while Figure \ref{fig_csss_hr_seq_b} demonstrates its capability to detect HR in a sitting-up position. We remark that the proposed approach cannot track rapid HR changes reflecting heart rate variability (HRV),  contributing to the discussed HR errors. We also observe from Figure \ref{fig_csss_hr_seq_a} that Soli fails to detect HR in the first half hours for this example.  Enhancing the recall rate of sleep tracking and investigating the feasibility of monitoring HRV using Soli technology could be of great interest.

\subsection*{Meditation Tracking}

The meditation data collection (DC) platform is a modification of the sleep-tracking platform, with the ECG sensor replaced by a fingertip PPG for convenience. To evaluate the performance of this approach, we conducted 24 test cases (outlined in Table \ref{tab_mediation}) with users in the supine position. Categories A, B, and C represent the primary cases with users lying still at two different distances ($0.6$ m and $1.0$ m) from the sensor to the chest, two blanket conditions (with and without blanket), and three breathing patterns (regular, deep, and rapid breathing). Users were asked to perform jumping jacks before DC for rapid breathing cases. Cases of Category D are used to test the approach's robustness against various body motions. Cases of Category E are used to test the approach's robustness against a second-person aggressor lying, standing, or walking around the DC environment. It is worth mentioning that the objective of this study is not to track the user's HR while in motion. Instead, the HR monitoring is temporarily halted when substantial body movements are detected by the presence detection block depicted in Figure \ref{fig1c} and resumed when the user is motionless again. As a result, the motion duration in Cases of Category D is not factored into the performance assessment. Additionally, this study does not aim to detect the HR of the second person in cases of Category E.

\begin{figure}
     \centering
     \begin{subfigure}[b]{0.5\textwidth}
         \centering
         \includegraphics[width=\textwidth]{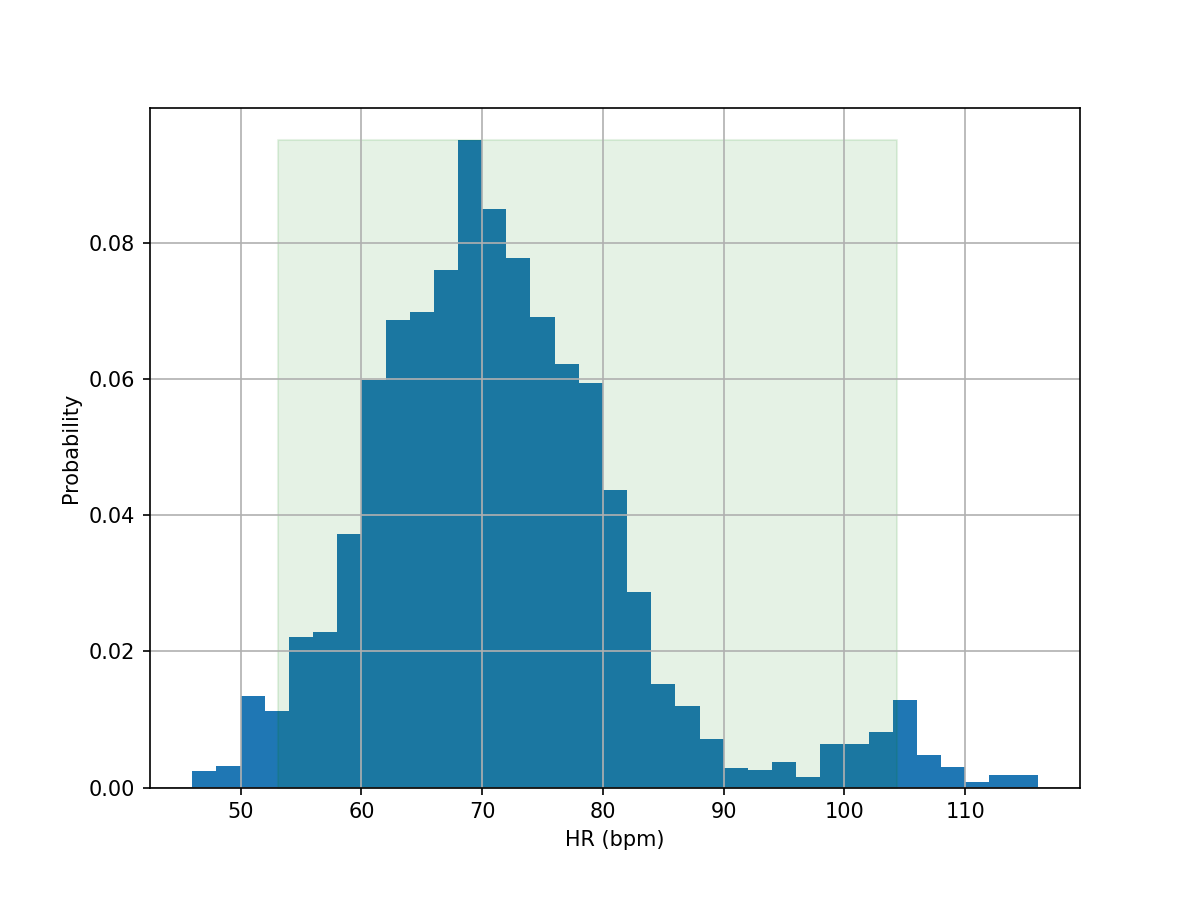}
     \end{subfigure}
     \caption{The meditation PPG HR histogram ranges from $45$ bpm to $115$ bpm.}
     \label{fig_meditation_hr_dist}
\end{figure}

In this meditation study, each data session lasts $2$ to $3$ minutes, with each user performing $5$ to $7$ sessions from different test cases. We have collected $1131$ minutes of data from 114 users, with the data sizes for each test case presented in Table \ref{tab_mediation}. Figure \ref{fig_meditation_hr_dist} displays the distribution of ground-truth HR obtained from the PPG data, which shows slightly higher HRs ($45$ to $115$ bpm) compared to its sleep counterpart (Figure \ref{fig_csss_dist_b}). It is important to note that a separate dataset has been collected for ML training for this meditation study. Thus, all data in Table  \ref{tab_mediation} were used for performance evaluation. {The data collection for ML training followed the same protocol outlined in Table \ref{tab_mediation}, albeit with a distinct set of subjects. We gathered approximately $876$ minutes of data from a cohort of $60$ participants.}

\begin{table}[ht]
\centering
\small
\begin{tabular}{|c|c|c|l|l|l|c|c|}
\hline
\textbf{Cases} & \textbf{Distance (m)} & \textbf{Blanket}&  \textbf{Breathing}  &  \textbf{Motion} &  \textbf{2nd person aggressor} &  \textbf{Users} &\textbf{Time (min.) }\\
\hline
A1 & 0.6 & No & Regular & Still & No &34& 84 \\
\hline
A2 & 0.6 & Yes & Regular & Still & No &33 & 72 \\
\hline
A3 & 1.0 & No & Regular & Still & No  &26& 52 \\
\hline
A4 & 1.0 & Yes & Regular & Still & No &36  & 88\\
\hline
B1 & 0.6 & No & Deep & Still & No  & 26&54\\
\hline
B2 & 0.6 & Yes & Deep & Still & No  & 29 & 58\\
\hline
B3 & 1.0 & No & Deep & Still & No  & 14 &30\\
\hline
B4 & 1.0 & Yes & Deep & Still & No  & 30&61\\
\hline
C1 & 0.6 & No & Rapid & Still & No  & 24&50\\
\hline
C2 & 0.6 & Yes & Rapid & Still & No & 31&63  \\
\hline
C3 & 1.0 & No & Rapid & Still & No  & 14&30\\
\hline
C4 & 1.0 & Yes & Rapid & Still & No  & 30&60 \\
\hline
D1 & 0.6 & No & Normal & Getting into bed & No& 31& 63 \\
\hline
D2 & 1.0 & No & Normal & Getting into bed & No  & 30& 60 \\
\hline
D3 & 1.0 & Yes & Rapid & Getting into bed & No  &9& 28 \\
\hline
D3 & 0.6 & No & Normal & Reaching device & No &9& 25 \\
\hline
D4 & 0.6 & No & Normal & Turning head & No  & 9&25 \\
\hline
D5 & 0.6 & No & Normal & Adjusting body & No  &35& 76 \\
\hline
D6 & 1.0 & No & Normal & Adjusting body & No &30 &61\\
\hline
D7 & 0.6 & No & Normal & Shifting location & No &30& 61  \\
\hline
E1 & 0.6 & No & Normal & Still & Walking along bed foot &2 & 6\\
\hline
E2 & 0.6 & No & Normal & Still & Lying still on non-device side&2 & 6\\
\hline
E3 & 0.6 & No & Normal & Still & Walking around device side & 2&6\\
\hline
E4 & 0.6 & No & Normal & Still & Getting into bed from non-device side& 2&  6\\
\hline
E5 & 0.6 & No & Normal & Still & Standing still behind device & 2&6\\
\hline
All & & &  & & & 114&1131 \\
\hline
\end{tabular}
\caption{The meditation test cases and related data size. 1131 minutes of data were collected for 24 test cases from 114 distinct users.}
\label{tab_mediation}
\end{table}

\begin{table}[ht]
\centering
\begin{tabular}{|p{0.1\linewidth}|p{0.1\linewidth}|p{0.1\linewidth}|p{0.1\linewidth}|p{0.1\linewidth}|p{0.1\linewidth}|p{0.1\linewidth}|}
\hline
\textbf{Method} & \textbf{MAE} & \textbf{AE} ($95\%$) &  \textbf{MAPE}  &  \textbf{APE} ($95\%$) & {\textbf{R$^2$}} &  \textbf{Recall} \\
\hline
\textbf{Proposed} & $\mathbf{1.05}$ bpm & $\mathbf{3.24}$ bpm & $\mathbf{1.56\%}$ & $\mathbf{4.69\%}$ &  { $\mathbf{95.26\%}$} & $\mathbf{98.16\%}$   \\  
\hline
\end{tabular}
\caption{The Soli HR accuracy of the meditation data. The proposed approach achieves MAE and MAPE around $1.05$ bpm and $1.56\%$, respectively. }
\label{tab_mediation_accu}
\end{table}

\begin{figure}
     \centering
     \begin{subfigure}[b]{0.45\textwidth}
         \centering
         \includegraphics[width=\textwidth]{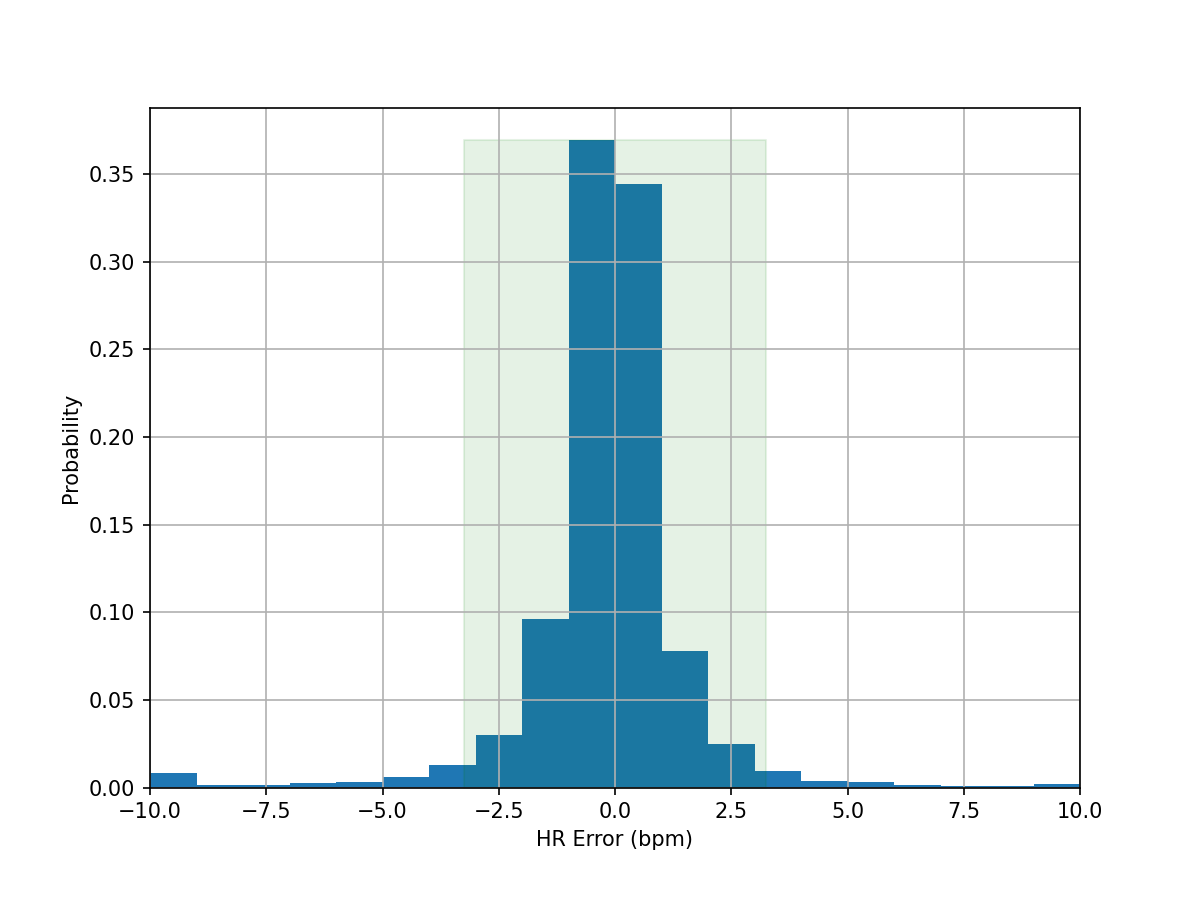}
         \caption{}
         \label{fig_meditation_dist_a}
     \end{subfigure}
     \begin{subfigure}[b]{0.45\textwidth}
         \centering
         \includegraphics[width=\textwidth]{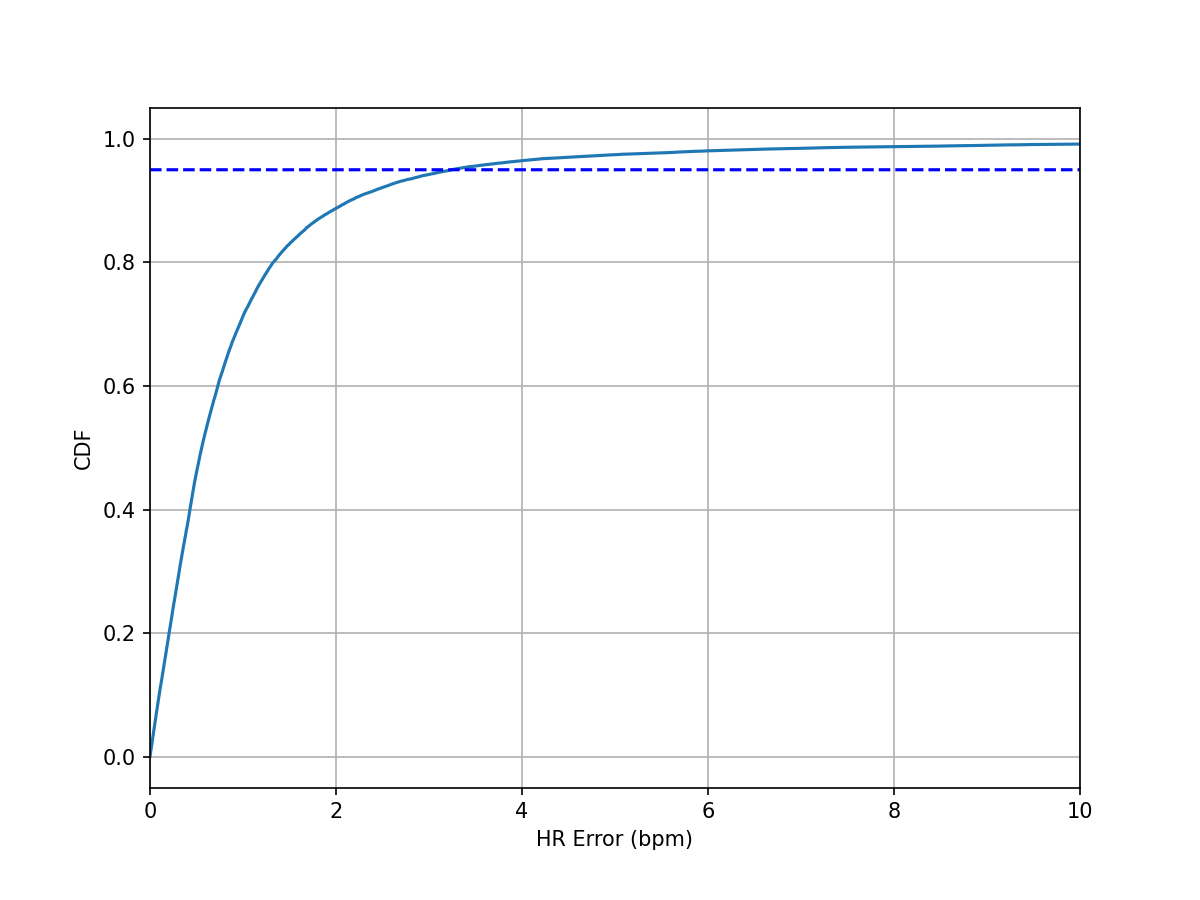}
         \caption{}
         \label{fig_meditation_dist_b}
     \end{subfigure}
     \caption{(a) HR error histogram of the proposed approach during meditation, and (b) the related AE CDF. The HR error histogram shows that most of the HR errors are small. The AE CDF shows that the $95$th percentile AE is less than $3.24$ bpm. }
     \label{fig_meditation_dist}
\end{figure}

\begin{figure}
     \centering
     \begin{subfigure}[b]{0.9\textwidth}
         \centering
         \includegraphics[width=\textwidth]{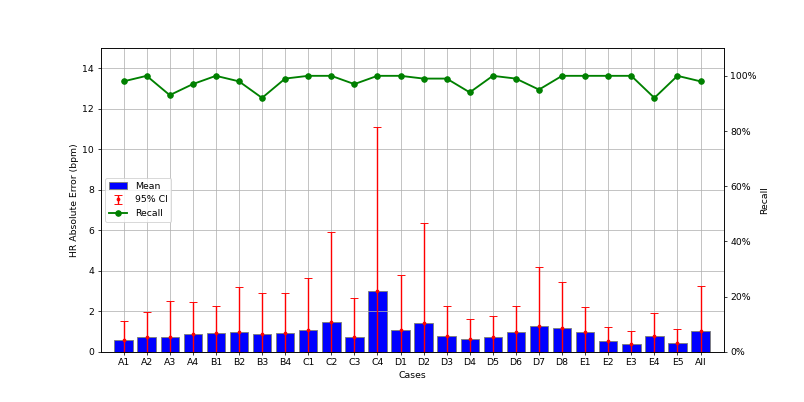}
     \end{subfigure}
     \caption{HR accuracy of the proposed method for various meditation test cases. The proposed approach achieves high accuracy across all cases, with the MAE of all cases except Case C4 less than $1.5$ bpm and the corresponding $95$th percentile AE less than $6.5$ bpm. }
     \label{fig_meditation_mae_vs_cases}
\end{figure}

\begin{figure}
     \centering
     \begin{subfigure}[b]{0.45\textwidth}
         \centering
         \includegraphics[width=\textwidth]{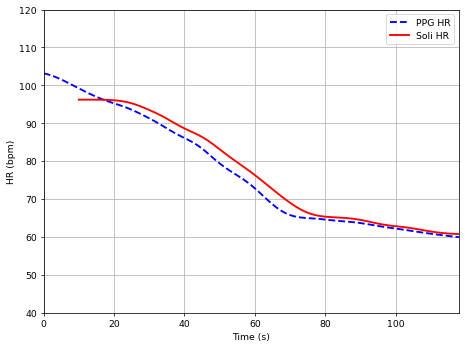}
         \caption{}
         \label{fig_meditation_example_a}
     \end{subfigure}
     \begin{subfigure}[b]{0.45\textwidth}
         \centering
         \includegraphics[width=\textwidth]{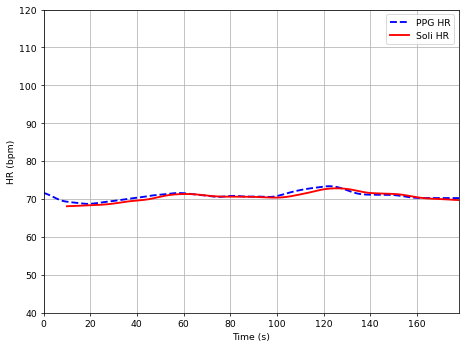}
         \caption{}
         \label{fig_meditation_example_b}
     \end{subfigure}
     \caption{Two meditation HR examples using our proposed approach. (a) shows an example in Case C3, where the user performed jumping jacks before data collection, and (b) shows an example in Case E2, where a second person lay still on the non-device side. In both cases, the Soli HR matches well with the PPG ground truth, indicating the accuracy of our proposed approach. }
     \label{fig_meditation_examples}
\end{figure}

For the meditation application, we employ the same SP algorithm and ML model utilized in sleep tracking. However, we make specific adjustments to the sample length and step size, setting them to $16$ and $4$ seconds, respectively, to ensure compliance with the latency requirements of the meditation application. Consequently, in this meditation study, our proposed approach delivers the initial heart rate measurement within a timeframe of $16$ seconds, followed by subsequent updates every $4$ seconds.

Table  \ref{tab_mediation_accu} presents the HR accuracy achieved by our proposed approach on the  meditation dataset, and Figures \ref{fig_meditation_dist_a} and \ref{fig_meditation_dist_b} show the related HR error histogram and CDF of HR AE, respectively. The proposed approach achieves good HR accuracy, with MAE and MAPE values of $1.05$ bpm and $1.56\%$, respectively, and the corresponding 95th percentiles of AE and APE being $3.24$ bpm and $4.69\%$. The recall rate is $98.16\%$, with only $1.84\%$ of samples remaining undetermined. {The R-Squared value is around $95.26\%$}. All of these performance metrics outperform their sleep counterparts. 

Figure \ref{fig_meditation_mae_vs_cases} presents the HR accuracies for various test cases. The proposed approach achieves high accuracy across all cases, with the MAE of all cases except Case C4 being less than $1.5$ bpm and the corresponding $95$th percentile AE being less than 6.5 bpm. Case C4 is slightly worse than the others, with MAE and the 95th percentile CI being $3.0$ bpm and $11.1$ bpm, respectively. Interestingly, Cases C4 and D3 perform differently even under similar conditions, possibly due to outliers. Overall, the figure shows the resilience of the proposed method, showcasing its ability to operate effectively up to 1.0 m with a blanket and successfully handle various breathing patterns, body motions, and even second-person aggressors (despite the limited data size of Category E). 

Figures \ref{fig_meditation_example_a} and  Figure \ref{fig_meditation_example_b} illustrate two examples of meditation in test cases C3 and E2, respectively. In Figures \ref{fig_meditation_example_a}, a decrease in HRs is observed from both the Soli and PPG sensors, which can be attributed to the jumping jack exercise performed before data collection. On the other hand, Figure \ref{fig_meditation_example_b}  depicts a scenario where a second person lies still on the non-device side.
In both cases, the Soli HRs exhibit strong alignment with the PPG HRs, indicating a high level of agreement between the two measurements. The MAEs for Figures \ref{fig_meditation_example_a}  is around $1.4$ bpm, while Figures \ref{fig_meditation_example_b} , it is $0.8$ bpm.

\section*{Discussion}

Detecting HR without physical contact is a valuable yet challenging task, especially for sleep and meditation tracking. This study presents a novel solution that leverages a miniaturized Soli radar chip integrated into a portable device (Google Nest Hub). The proposed approach harnesses advanced SP and ML techniques to extract HR information from radar signals, enabling precise measurements during sleep and meditation sessions. The effectiveness of the proposed approach is validated using two datasets: a sleep dataset comprising data from 62 users and 498 hours and a meditation dataset with data from 114 users and 1131 minutes. The results show that the proposed approach achieves an MAE of 1.69 bpm and a MAPE of $2.67\%$ on the sleep dataset and an MAE of 1.05 bpm and a MAPE of $1.56\%$ on the meditation dataset. The recall rates for the two datasets are $88.53\%$ and $98.16\%$, respectively.  Furthermore, the proposed method is robust against various sleep positions, HR bands, body motions, and even the presence of a second-person aggressor. These results suggest that noncontact HR monitoring technology has promising potential for continuous and convenient monitoring of sleep and meditation.

Future work will focus on enhancing the sleep-tracking recall rate and improving Soli's capability to track rapid HR changes for HRV detection, another critical biometric for meditation. Another interesting research area is RF-Seismocardiogram (SCG), which aims to extract more comprehensive heartbeat information beyond HR and HRV for monitoring various cardiovascular conditions of the user, especially during sleep\cite{HaAssanaAdib2020}. With further developments, noncontact heart monitoring technology enabled by Soli and other radar technologies may become essential for continuously and conveniently monitoring sleep and meditation.

\section*{Methods}

\subsection*{Hardware}

The second-generation Soli radar chip, illustrated in Figure \ref{fig1b}, is utilized in our investigation. The chip is integrated into the Google Nest Hub, as depicted in Figure \ref{fig1a}. This compact chip measures  $6.5 \mbox{ mm} \times 5 \mbox {mm} \times 0.9 \mbox{ mm}$ and operates at $60$ GHz with one transmit antenna and three receive antennas. Its diminutive size allows for integration into various portable devices and smartphones. The receive antennas are configured in an L shape, with $2.5$ mm spacing between them. The chip employs frequency-modulated continuous wave (FMCW) waveforms \cite{AlizadehShakerAlmeida2019}, also known as chirps, with a transmit power of $5$ mW. These chirps sweep frequencies from $58$ GHz to $63.5$ GHz, resulting in a bandwidth of $B = 5.5$ GHz and a range resolution of approximately $\frac{c}{2B}= 2.7$ cm, where $c$ denotes the speed of light. The received signals are sampled at an ADC sampling rate of $2$ MHz, with each chirp comprising $256$ samples. Chirps are repeated at a rate of $3000$ Hz, with $20$ chirps organized into a burst that repeats at a burst rate of $30$ Hz. The chip enters an idle mode between chirps and bursts, resulting in an active duty cycle of roughly $\frac{256*20}{2\times10^6} *30 = 7.68\%$ within $33$ ms, compliant with the FCC wavier \cite{fcc_da_18_1308a1}. The essential operational parameters of the Soli radar chip are presented in Table \ref{tab_soli_parameters}.  For a more comprehensive understanding of the Soli radar chip and the FMCW principle, interested readers may refer to the related papers \cite{TrottaWeber2021, HayashiLien2021,LienGuillianKaragozler2016,Mercuri2019}.

\begin{table}[ht]
\centering
\begin{tabular}{|p{0.3\linewidth}|p{0.2\linewidth}|}
\hline
\textbf{Parameters} & \textbf{Values} \\ 
\hline
Frequency band &  $58\textup{--}63.5$ GHz \\
\hline
Transmit power & 5 mw\\
\hline
ADC sampling rate & 2 MHz \\
\hline
Number of samples per chirp & 256 \\
\hline
Number of chirps per burst & 20 \\
\hline
Chirp rate & 3000 Hz\\
\hline
Burst rate & 30 Hz \\
\hline
Active duty cycle & $7.68\%$ \\
\hline
\end{tabular}
\caption{Soli radar chip operation parameters. Low transmit power ($5$ mW) and active duty cycle ($7.58\%$) were used to meet the FCC waiver.}
\label{tab_soli_parameters}
\end{table}

\begin{figure}
     \centering
     \begin{subfigure}[b]{0.45\textwidth}
         \centering
         \includegraphics[width=\textwidth]{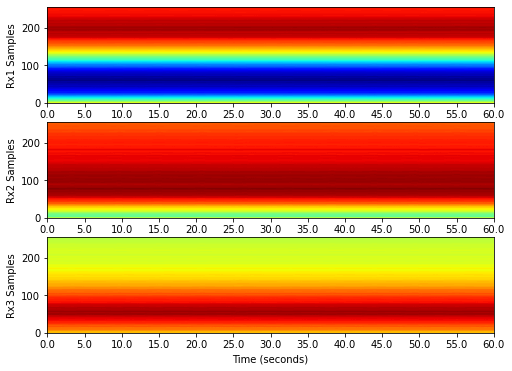}
         \caption{}
         \label{fig_csss_data_sample_a}
     \end{subfigure}
     \begin{subfigure}[b]{0.45\textwidth}
         \centering
         \includegraphics[width=\textwidth]{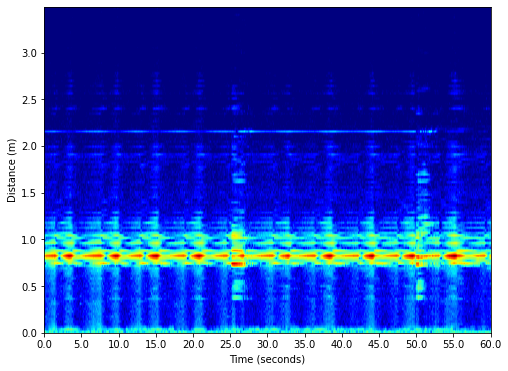}
         \caption{}
         \label{fig_csss_data_sample_b}
     \end{subfigure}
     \medskip
     \begin{subfigure}[b]{0.45\textwidth}
         \centering
         \includegraphics[width=\textwidth]{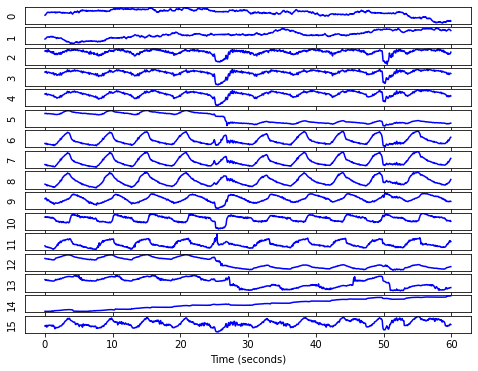}
         \caption{}
         \label{fig_csss_data_sample_c}
     \end{subfigure}
     \medskip
     \begin{subfigure}[b]{0.45\textwidth}
         \centering
         \includegraphics[width=\textwidth]{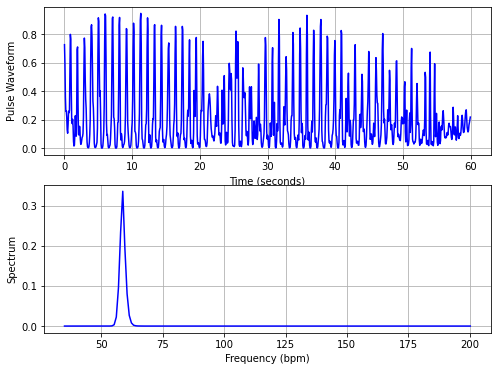}
         \caption{}
         \label{fig_csss_data_sample_d}
     \end{subfigure}
        \caption{An example of noncontact HR detection for sleep tracking includes (a) ADC samples over 60 seconds of the three receivers, (b) combined power range profiles for presence detection, (c) extracted 16 micro-motion waveforms, and (d) pulse waveform and associated pseudo-spectrum at the NN output. }
        \label{fig_csss_data_sample}
\end{figure}

\subsection*{Preprocessing}

The ADC samples from the three receivers are initially directed to the preprocessing block, as shown in Figure \ref{fig1c}. Within the preprocessing block, the chirps within each burst are first averaged. Then the bursts are decimated from 30 Hz to $15$ Hz by merely averaging two adjacent bursts, enhancing the SNR. Consequently, the data is decimated to one chirp every $0.067$ seconds. FFT is then applied to the decimated data along the fast-time, i.e., the $256$ samples of each chirp, generating complex-valued range profiles. An example of decimated data from the three receivers of $60$-second duration is illustrated in Figure \ref{fig_csss_data_sample_a}, with the x-axis representing time and the y-axis representing the ADC sample index with a chirp.

\subsection*{Presence Detection}

The presence detection block in this study serves three functions: firstly, to determine whether a user is present; secondly, to determine whether the user is still; and thirdly, to determine the distance of the user from the sensor. When the presence detection block detects a still user, it will activate the subsequent HR monitoring blocks to measure the user's HR. Conversely, it pauses the HR detection process until a still user is detected again. By including these functionalities in the presence detection block, the system can optimize HR monitoring, ensuring accurate and reliable readings are obtained. 

In the presence detection stage, we first apply a clutter filter to complex-valued range profiles to eliminate strong stationary clutter originating from the background and the interplay between transmit and receive antennas. The clutter filter first calculates the average range profile over time (or chirps) and subtracts it from the original ones. It assumes that these clutter signals remain unvaried over time. The power of the so-processed range profiles is then computed and combined over receivers to form a  power range profile image. As shown in Figure \ref{fig_csss_data_sample_b}, the bright line in the power range profile image signifies user presence. A simple constant false alarm (CFAR) \cite{Richards2005} detector is used to detect  the user and determine the associated distance. Upon user detection, the peak range bin signal is extracted from the complex-valued range profiles (after the clutter filter). An FFT is then executed to obtain a Doppler spectrum. Given that Doppler indicates the user's motion speed, the ratio of low Doppler energy to high Doppler energy is calculated to determine the user's stillness.

\subsection*{Micro-motion Extraction}

Once a still user  is detected, $16$ range bin signals are extracted from the complex-valued range profiles around the detected user distance. The micro-motion extraction algorithm is applied to each range bin independently, resulting in a micro-motion waveform. 

The micro-motion extraction process involves two main steps: beamforming and phase extraction. In the first step, signals from the three receivers are combined using a maximum ratio combining (MRC) method to filter out the desired reflection signal. To perform the MRC beamforming, we first stack the signals from the three receivers into a $3 \times 1$ vector, denoted by $\xbf(l) \in \mathcal{C}^{3\times 1}$, where $l$ denotes the chirp index. A $3 \times 3$ covariance matrix $\Qbf$ is then computed using the formula: 
\begin{align}
&&\Qbf  =  \frac{1}{L}\sum_{t=1}^L \left[ \xbf(l) - \bar \xbf\right] \left[ \xbf(l) - \bar \xbf \right]^H.
\label{equ_q}
\end{align}
In (\ref{equ_q}), $\hat \xbf = \frac{1}{L} \sum_{l=1}^L \xbf(l)$ represents the stationary clutter at the current range bin, $(\cdot)^H$ denotes the Hermitian matrix transpose, and $L$ is the number of chirps (in the example in Figure \ref{fig_csss_data_sample}, this is $900$). The dominant eigenvector of $\Qbf$, denoted by $\wbf$, is computed using a power iteration method \cite{GolubVanLoan1996}. Finally, the three channels are combined using $\wbf$ to produce a complex-valued scalar sequence $y(l) = \wbf^H \xbf(l)$. 

The objective of the second step is to extract phase information from the combined signal  $y(l)$. Note that the conventional phase extraction technique is ineffective in this problem due to the small heartbeat and respiration micro-motion magnitude compared to the wavelength. Instead, a circle-fitting technique  \cite{AlizadehShakerAlmeida2019, mercuri2022automatic} is used. This technique assumes that the desired reflection signal has a constant modulus over time. Thus, it can be formulated in the following optimization problem: 
\begin{align}
&&\min_{\eta, r} \sum_{l} \left[ \left| y(l) - \eta \right|^2 - r \right]^2,
\label{equ_circle_fitting}
\end{align}
where $\eta \in \mathcal{C}$ and $r \in \mathcal{R}$ represent the center and radius of the fitted circle, respectively.  The optimization problem of (\ref{equ_circle_fitting}) can be solved using a closed-form solution \cite{AlizadehShakerAlmeida2019}.  
 
Once we estimate the circle center $\eta$, we can extract the micro-motion waveform using the following equation: 
\begin{align}
&&d(l) = \frac{c}{4\pi f_0} \mbox{unwrap}\left[ \mbox{angle} (y(l) - \eta) \right],
\label{equ_d}
\end{align}
with $f_0$ denoting the center frequency. Here, $\mbox{unwrap}(\cdot)$ denotes the unwrap function, which corrects the radian phase angles by adding multiples of $\pm 2 \pi$  such that the phase change from the previous sample is less than $\pi$.

Figure \ref{fig_csss_data_sample_c} displays the $16$ micro-motion waveforms extracted from the sleep  data example. While some waveforms exhibit respiration motion, the heart pulse signal is scarcely discernible. We employ an ML technique below to extract the pulse waveform and its associated pseudo-spectrum from the micro-motion waveforms.

\subsection*{ML-based Pulse Waveform and Pseudo-Spectrum Extraction}

\begin{figure}
     \centering
     \begin{subfigure}[b]{1\textwidth}
         \centering
         \includegraphics[width=\textwidth]{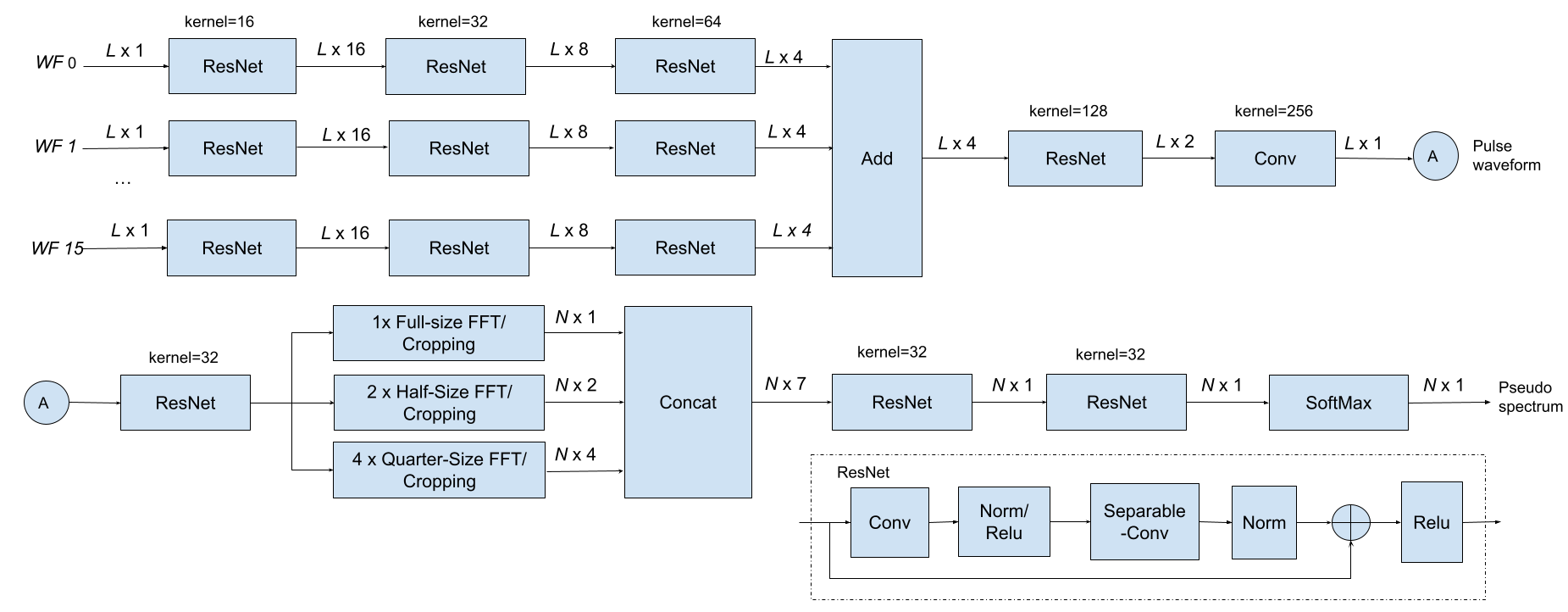}
     \end{subfigure}
     \caption{A lightweight NN for HR detection. The NN consists of two blocks. The first block extracts the pulse waveform from the 16 micro-motion waveforms, while the second block generates a pseudo-spectrum from the pulse waveform. The designed NN has 10,248 parameters, 9,815 of which are trainable. }
     \label{fig_hr_ml_model}
\end{figure}
 
A neural network (NN) has been designed to extract the heart pulse waveform and its corresponding pseudo-spectrum from the micro-motion waveforms. The NN comprises two blocks, depicted in Figure \ref{fig_hr_ml_model}.  The first block focuses on extracting the pulse waveform from the $16$ micro-motion waveforms, as shown in Figure \ref{fig_csss_data_sample_c}. The so-obtained pulse waveform is illustrated in the upper plot of Figure \ref{fig_csss_data_sample_d}. The second block takes this pulse waveform as input and generates a pseudo-spectrum, visualized in the lower plot of Figure \ref{fig_csss_data_sample_d}.

At the bottom of Figure \ref{fig_hr_ml_model}, we present a modified residual neural network (ResNet) layer \cite{He_2016_CVPR}, which is utilized in both neural network (NN) blocks \cite{He_2016_CVPR}.
The modified ResNet layer comprises a 1D convolution (Conv) layer, followed by a separable convolution (Separable-Conv)\cite{Chollet_2017_CVPR} layer. We use Separable-Conv for the second stage to reduce the number of parameters. A shortcut,  two batch normalizations, and two rectified linear units (ReLu) are included,  similar to the standard ResNet. 

As shown in Figure \ref{fig_hr_ml_model}, in the first block,  each of the $16$ micro-motion waveforms is processed through three ResNet layers before being passed to a summation layer.  The output of the summation layer is then processed through a ResNet and a standard 1D Conv layer, ultimately producing the pulse waveform output. The number of filters and their corresponding kernel size for each ResNet layer is illustrated in the figure. The ResNet weights across various input branches are enforced to be equal.

The second  NN block begins by processing the pulse waveform through a ResNet layer. The resulting output is then passed through a full-size FFT layer, two half-size FFT layers, and four quarter-size FFT layers in parallel. In the half-size FFT layers, the input signal is divided into two segments, and FFT is applied to each segment independently. A similar process is used for the quarter-size FFT block. This design enhances the robustness of HR detection against random body motion, which is usually non-stationary and thus with  different spectra at different time intervals. By examining the spectra at different FFT outputs, the NN can better reject the interference of random body motion. Zero-padding is applied to all $7$ FFTs resulting in the same output length ($1024$) and the same output frequency granularity ($0.88$ bpm). The FFT outputs are then cropped to the frequency band of interest ($35\textup{--}200$ bpm, $N=189$) and concatenated. The concatenated data is processed by two ResNet layers and one SoftMax layer, generating a pseudo-spectrum of the pulse waveform. The HR and associated confidence level can be determined by detecting the peak of the pseudo-spectrum and comparing its amplitude with that of the second peak.

{The designed NN is very lightweight, with only 10,248 parameters, of which 9,815 are trainable. As a result, the entire pipeline, including the NN model, data acquisition/transmission, signal processing, and virtualization, can run in real time on a laptop (e.g., MacBook Pro with 2.6 GHz 6-Core Intel Core i7) or a portable device (e.g., Google Nest Hub with Amlogic S905D3 quad-core Cortex-A55 processor), updating the HR estimate every 4 seconds.}
The NN was implemented using TensorFlow 2.13.0 \cite{tensorflow2015-whitepaper, abadi2016tensorflow} and trained in the Google Colaboratory environment \cite{CarneiroMedeiros2018}. We used $477$ hours of sleep data and $478$ minutes of meditation data for training. The training data were collected from different users using the same DC protocols as the test data. The neural network (NN) model was trained separately for sleep and meditation tracking applications due to the differences in input data lengths. 

During the NN training, we utilized both the ECG and PPG waveforms, along with their respective pseudo-spectra, as ground truth labels. To facilitate ML training, we pre-processed and normalized both the ECG/PCG waveforms and their pseudo-spectra. Initially, we detected individual peaks from the raw ECG/PCG waveforms, which were then modulated by Gaussian pulses. This process effectively replaced the heartbeat pulses with standard Gaussian pulses, preserving the timing information of the heartbeats while removing other unrelated features. A similar approach was applied to process the spectrum labels. For the waveform output, a mean square error (MSE) loss function was employed, while a cross-entropy loss function was used for the pseudo-spectrum output. By combining these two loss functions, we optimized the overall performance of the NN. The Adam optimizer \cite{goodfellow2016deep} was utilized to facilitate the optimization process.

\subsection*{Post-Processing}

The HR sequence obtained from the signal processing (SP) and machine learning (ML) blocks undergoes further refinement in the post-processing block. Initially, HRs with low confidence levels are discarded, and a simple linear interpolation technique is employed to recover these rejected HR samples using neighboring samples as reference. However, if the consecutive number of rejected samples exceeds the subsequent median filter length, the HRs during that period are considered as undetermined.

A median filter is applied to ensure smoothness in the HR sequence, followed by a Gaussian smoothing filter. For sleep tracking, the median filter length is set to $10$ minutes, while the Gaussian smoothing filter has a length of $1$ minute. The median and Gaussian filter lengths for meditation tracking are set to $20$ seconds. These filter lengths are chosen to optimize the balance between preserving accurate HR information and reducing noise in the respective tracking scenarios.

\subsection*{Performance Metrics}

The accuracy of Soli HR is evaluated using five performance metrics: recall rate, MAE, MAPE, 95th percentile AE, and 95th percentile APE. The recall rate measures the percentage of data samples with determined HR values. It is calculated by dividing the number of samples with determined HRs by the total number of samples. 
AE and APE are computed for each determined HR sample to quantify the differences between the Soli HR and the ground-truth HR. The MAE is then determined by averaging the AEs over the investigated sample set, providing a measure of the average absolute error, i.e., 
\begin{align}
&& \mbox{MAE} = \frac{1}{N} \sum_{n=1}^N |\mbox{hr}_n - \widehat{\mbox{hr}}_n|,
\end{align}
with $\mbox{hr}_n$ and $\widehat{\mbox{hr}}_n$ representing the reference and estimated HRs, respectively. Similarly, the MAPE is calculated by averaging the APEs, indicating the average percentage error,  i.e.,
\begin{align}
&& \mbox{MAPE} = \frac{1}{N} \sum_{n=1}^N \frac{|\mbox{hr}_n - \widehat{\mbox{hr}}_n|}{\mbox{hr}_n}.
\end{align}
 Additionally, the 95th percentile AE and 95th percentile APE are calculated from the distribution of AEs and APEs, respectively. These metrics represent the value below which $95\%$ of the absolute or percentage errors fall, providing insight into the performance at higher error levels. 

{In Tables \ref{tab_sleep_sense_hr_accu} and \ref{tab_mediation_accu}, we have also included the R-squared value, denoted as $\mbox{R}^2$, as an additional metric to offer insights into the model's performance. This statistical measure quantifies the proportion of HR variation that the model can predict compared to the total variation in HR. It is calculated as:
\begin{align}
&& \mbox{R}^2 = 1 - \frac{ \sum_{n=1}^N  |\mbox{hr}_n - \widehat{\mbox{hr}}_n|^2}{ \sum_{n=1}^N 
|\mbox{hr}_n - \mywidebar{\mbox{hr}}|^2},
\label{equ_r_square}
\end{align}
with $\mywidebar{\mbox{hr}} = \frac{1}{N} \sum_{n=1}^N {\mbox{hr}_n}$ representing the averaged HR over all data samples.

\section*{Ethical Statement} All experiments and methods were performed in accordance with the relevant guidelines and regulations. Informed consent was obtained from all subjects and/or their legal guardian(s). The human heart rate collection protocol was approved by the Institutional Review Board of Google (IRB no. Pro00042166). 

\section*{Data availability} The datasets produced and/or analyzed during the study are currently not publicly available. Requests for access to the data may be directed to the corresponding author.

\bibliography{vital}

\section*{Acknowledgments} The authors gratefully acknowledge the support and contributions of Jessie Yang, Keith Klumb, Jenna Drumright, Billie Whitehouse, Madeleine Gong, Ben Wong, Alex Paniutin, Ranjit Deshpande, Shruti Gupta, Alan Laursen,  William Tran, Rafal Protasiuk, Farhana Khan, Andrew Bartunek and other unnamed individuals in program management, hardware and software development, and data collection. 

{
\section*{Competing interests}
This study was funded by Alphabeta Inc. All authors are affiliated with Alphabet, either as current employees or former employees, and may hold stock as a component of their standard compensation package.
}








\end{document}